\documentclass{new_tlp}
\usepackage{url}
\usepackage{amssymb}

\usepackage[cached,lncs]{FormulaLangInt}

\newcommand{\frm}{{\sc formula}}
\newcommand{\dc}{\_\!\_}
\newcommand{\head}{\textrm{~:-~}}
\newcommand{\dct}{\rule{1.5mm}{.1pt}}
\newcommand{\symbols}{\ensuremath{\mathcal{S}}}
\newcommand{\kinds}{\ensuremath{\mathcal{K}}}
\newcommand{\terms}{\ensuremath{\mathcal{T}}}
\newcommand{\naturals}{\ensuremath{\mathcal{N}}}
\newcommand{\dtb}{\ensuremath{\stackrel{de\!f}{=}}}
\newcommand{\code}[1]{{\small\texttt{#1}}}
\newcommand{\keyw}[1]{{\color{formula-keyword}{#1}}}
\newcommand{\relabel}[2]{{\ensuremath{\rho_{#1\rightarrow #2}}}}

\newenvironment{CodeBlock}
  {\begin{flushleft}\ttfamily\small}
  {\end{flushleft}}

  \title[Theory and Practice of Logic Programming]
        {A Module System for Domain-Specific Languages}

  \author[E. K. Jackson]
         {ETHAN K. JACKSON\\
         Research In Software Engineering (RiSE)\\
         Microsoft Research, Redmond, WA, USA 98052\\
         \email{ejackson@microsoft.com}}

\submitted{}
\accepted{}
\revised{}

\begin{document}

\maketitle

  \begin{abstract}
\textit{Domain-specific languages} (DSLs) are routinely created to simplify difficult or specialized programming tasks. They expose useful abstractions and design patterns in the form of language constructs, provide static semantics to eagerly detect misuse of these constructs, and dynamic semantics to completely define how language constructs interact. However, implementing and composing DSLs is a non-trivial task, and there is a lack of tools and techniques. 

We address this problem by presenting a complete module system over LP for DSL construction, reuse, and composition. LP is already useful for DSL design, because it supports executable language specifications using notations familiar to language designers. We extend LP with a module system that is simple (with a few concepts), succinct (for key DSL specification scenarios), and composable (on the level of languages, compilers, and programs). These design choices reflect our use of LP for industrial DSL design. Our module system has been implemented in the \frm~language, and was used to build key Windows 8 device drivers via DSLs. Though we present our module system as it actually appears in our \frm~language, our emphasis is on concepts adaptable to other LP languages. 
\end{abstract}

  \begin{keywords}
    module systems, domain-specific languages, logic programming
  \end{keywords}

\section{Introduction}
\textit{Domain-specific languages} (DSLs) are routinely created to simplify difficult or specialized programming tasks \cite{CarteyLM12,VincentelliSSYM09}. They expose useful abstractions and design patterns in the form of language constructs. Unlike \textit{libraries}, which also expose abstractions, DSLs have static semantics to eagerly detect misuse of constructs and dynamic semantics to completely define how language constructs interact (independently of implementation). However, these advantages come with at least two disadvantages: (1) Language design and implementation is challenging, requiring formal specifications, compilers, and debuggers. (2) Composing and reusing DSLs is non-trivial, whereas, on the surface, composing and reusing libraries is as simple as importing them and calling their APIs. 

\textit{Logic programming} (LP) is useful for DSL design and implementation \cite{Gurevich12,AlvaroMCHMS10}. Traditionally, programming languages have been specified with a combination of \textit{algebraic data types} (for ASTs) \cite{Hudak}, \textit{rules of inference} (for typing and static semantics) \cite{Cardelli97}, and \textit{abstract transition systems} (for dynamic semantics) \cite{Borger05}. These specification styles are closely related to logic programs, and rephrasing them as proper logic programs yields formal specifications of DSL semantics that are also language implementations. However, existing LP systems have much less support for composing and reusing DSL specifications. Most systems provide modules for defining and exporting predicates, but this is a low-level form of composition from the perspective of DSLs and compilers \cite{HaemmerleF06}. 

In this paper we present a complete module system over LP for DSL construction, reuse, and composition. Table \ref{tab:modsys} gives an overview of our module system and the structure of this paper.  It has been designed to be simple (with a few concepts), succinct (for key DSL specification scenarios), and composable (on the level of languages, compilers, and programs). These design choices reflect our use of LP for industrial DSL design. Our module system has been implemented in the \frm~language (\url{http://formula.codeplex.com}), and has been used to build DSLs for modeling cyber-physical systems \cite{SimkoLLNS13}, programming Windows device drivers \cite{DesaiGJQRZ13}, and specifying resource allocation problems \cite{JacksonKDSS10}, to name a few. Windows 8.0 ships with core components that were built using DSLs specified with \frm~(e.g. the USB 3.0 stack).

\begin{table}[t]
\begin{tabular}{p{0.13\textwidth} p{0.7\textwidth} p{0.1\textwidth}}
\multicolumn{3}{c}{\hspace{-7pt}\fcolorbox{formula-background-gray}{formula-background-gray}{
\begin{minipage}{0.975\textwidth}
 \begin{center}
 \textbf{Modules}
 \end{center}
\end{minipage}}}\\[-5pt] \hline \textbf{Kind} & \textbf{Purpose} & \textbf{Section}\\[-5pt] \hline
Domain & Describes DSL syntax, type judgments, and static semantics using \textit{algebraic data types} (ADTs) and LP.\vspace{4pt} & \ref{sec:syntax}, \ref{sec:judge} \\
Model & Represents a DSL program w.r.t. a domain as a set of ground facts.\vspace{4pt} & \ref{sec:models}\\ 
Transform & A function from models to models, such as a compiler or transition system. Defined using ADTs and LP.  & \ref{sec:transforms} \\ 
\multicolumn{3}{c}{\hspace{-7pt}\fcolorbox{formula-background-gray}{formula-background-gray}{
\begin{minipage}{0.975\textwidth}
 \begin{center}
 \textbf{Language Features}
 \end{center}
\end{minipage}}}\\[-5pt] \hline \textbf{Feature} & \textbf{Purpose} & \textbf{Section}\\[-5pt] \hline
Contracts & \textit{Conforms clauses} specify DSL static semantics. \textit{Requires} and \textit{ensures} clauses specify transform behavior. \vspace{4pt} & \ref{sec:judge}, \ref{sec:exe} \\
Symbolic constants & Module-level constants for labeling common sub-expressions and naming value-level transform parameters.\vspace{4pt} & \ref{sec:models} \\ 
Inferred rewrites & Compiler-inferred term rewrites that can replace boilerplate recursive rules in transforms. & \ref{sec:rewrites} \\ 
\multicolumn{3}{c}{\hspace{-7pt}\fcolorbox{formula-background-gray}{formula-background-gray}{
\begin{minipage}{0.975\textwidth}
 \begin{center}
 \textbf{Composition Operators}
 \end{center}
\end{minipage}}}\\[-5pt] \hline \textbf{Feature} & \textbf{Purpose} & \textbf{Section}\\[-5pt] \hline
Extends,\hspace{5em}Includes & Compose DSL syntax and static semantics (for domains) and DSL programs (for models); \textit{extends} is conjunctive for static semantics. \vspace{4pt} & \ref{sec:domcomp}, \ref{sec:modelcomp} \\
Renaming (::) & Generates new modules by systematically renaming data constructors. Used to create product DSLs and define transforms.\vspace{4pt} & \ref{sec:renaming}  \\ 
Sequential composition & Creates larger transforms by sequential composition. For instance, a compiler can be decomposed into a sequence of small transforms. & \ref{sec:exe} \\ 
\hline
\end{tabular}
\caption{Overview of module system.}
\label{tab:modsys}
\end{table}

We demonstrate our module system by modularly developing a finite state machine DSL with a small action language. Our examples are as they appear in our \frm~language, but we emphasize the concepts adaptable to other LP languages. Related work is presented throughout the paper. 

\section{Domains, Models, and Conformance}
\begin{figure}[t]
\begin{FormulaCode}
:domain NonDetFSM {
:   // FSM Syntax
exISyntaxS:   State    ::= new (id: Integer).
:   Event    ::= new (id: String).
:   Trans    ::= new (src: State, ev: Event, dst: State).
exISyntaxE:   Init     ::= new (st: State).   
:   // Reachability judgment
exIJudgeS:   Reach    ::= (State).
exIJudgeE:   Reach(s) :- Init(s); Reach(s'), Trans(s', _, s).
:   // There must be an initial state.
exIConformsS:   conforms Init(_).
:   // Initial states must be "defined".
exIInitDef:   conforms no { i | i is Init, no { s | s is State, s = i.st } }.
:   // Transitions must be over "defined" states / events.
:   conforms no { t | t is Trans, no { s | s is State, s = t.src } }.
:   conforms no { t | t is Trans, no { s | s is State, s = t.dst } }.
exIConformsE:   conforms no { t | t is Trans, no { s | s is Event, s = t.ev  } }.             
:}
\end{FormulaCode}
\caption{Domain module defining a class of non-deterministic finite state machines.}
\label{fig:NonDetFSM}
\end{figure}

\textit{Domains} specify DSL syntaxes and static semantics, whereas \textit{models} represent DSL programs. There is a \textit{conformance} relationship between domains and models: a model conforms to its domain if it satisfies the domain's static semantics. Domains encapsulate \textit{algebraic data type} (ADT) definitions, rules of inference, and static semantics. Figure \ref{fig:NonDetFSM} shows the \textit{NonDetFSM} domain for a class of non-deterministic finite state machines. Lines \exISyntaxS~-~\exISyntaxE~are ADT definitions for the syntax of the DSL. Lines \exIJudgeS~and~\exIJudgeE~define the syntax and semantics of the \textit{Reach} judgment, which judges the reachable machine states. Finally, lines \exIConformsS~-~\exIConformsE~give the static semantics.

\subsection{Syntaxes and Algebraic Data Types}
\label{sec:syntax}
ADTs are standard for representing the syntaxes of DSLs and judgments, and our module system makes heavy use of them. Our ADT definitions come in two forms. The first form defines a \textit{data constructor} $F$:
\begin{displaymath}
F ::= [\mathbf{new}] \ ([arg_1: ] \ T_1, \ldots, [arg_n: ] \ T_n).\\
\end{displaymath}
where $[e]$ is an optional expression. This form introduces an $n$-ary data constructor $F()$ and an identically named type $F$. The type $F$ denotes all values obtained by applying $F()$ to arguments $a_1, \ldots, a_n$ belonging to types $T_1, \ldots, T_n$. For instance, the $State$ type denotes the set of all values constructed by applying $State()$ to an integral value. Applying $State()$ to a non-integral value creates a badly-typed value. The \frm~compiler statically guarantees that badly-typed values can never be created \cite{JacksonSB12}. The arguments of data constructors can optionally be named, e.g. the first argument of $State()$ is called $id$. Data constructors marked with the $\mathbf{new}$ keyword are used for DSL syntax, whereas unmarked constructors are used for the syntax of judgments. Notice that the $Reach()$ constructor is unmarked, and can never appear as part of DSL syntax.  The second form of type definition defines a \textit{union type} and has the form:
$U ::= T_1 + \ldots + T_n$.
Here, $U$ denotes the mathematical union of the types $T_1, \ldots, T_n$.

\textbf{Related work.}~~There are many variations of ADTs. Our ADTs are a sub-class of \textit{regular types}, with the useful properties that type equality is co-NP-complete and type expressions have canonical forms \cite{JacksonBS11}. Generalizing to full regular types significantly changes the complexity of static checking, as type equality becomes EXPTIME-complete and types become as expressive as \textit{tree automata} \cite{ImplRegTypes}. For instance, let $s()$ be a unary data constructor standing for the successor function, then regular types can distinguish between even and odd naturals with the productions:
\begin{displaymath}
Even \rightarrow 0. \quad Even \rightarrow s(Odd). \quad Odd \rightarrow s(Even).
\end{displaymath}
Few existing LP systems implement full regular types (e.g. \textit{Ciao} \cite{HermenegildoPBL05}). In our experience, a weaker type system is sufficient for capturing syntax, and then more complex properties can be stated directly using LP.

ADTs are also standard in functional programming languages where they are generalized along different dimension compared to regular types \cite{SulzmannWS06}. For example, consider these recursive type definitions for two different kinds of lists in the \textit{Haskell} language \cite{HeerenLI03}:
\begin{CodeBlock}
\keyw{data} ListInt = ConsInt  Integer  ListInt  |  NilInt\\
\keyw{data} ListStr = ConsStr  String~  ListStr  |  NilStr
\end{CodeBlock}
In Haskell each definition requires a distinct $Nil$ constructor (i.e. $NilI\!nt$ and $NilStr$), otherwise the type of $Nil$ would be both $ListI\!nt$ and $ListStr$. In our experience, this restriction is too strong, because it hinders composition of DSL syntaxes. In \frm~these two definitions can share the same $Nil$ constructor as follows:
\begin{CodeBlock}
ConsInt ::= (Integer, ListInt). ListInt ::= ConsInt + \{ Nil \}.\\
ConsStr ::= (String, ~ListStr). ListStr ::= ConsStr + \{ Nil \}.
\end{CodeBlock}
(The \frm~syntax $\{ c_1, \ldots, c_n \}$ denotes a set of constants.) Consequently, \frm~values do not have unique minimal types. This allows syntactic elements to be mixed more freely in composite languages, but comes at the cost of higher-complexity type equality and sub-type testing.

\subsection{Judgments and Conformance}
\label{sec:judge}
LP is used to define rules of inference and static semantics. For example, the reachability judgment (line \exIJudgeE) is a standard recursive rule marking a state as reachable if it can be reached from an initial state by some transitions. To be clear, \frm~does not support definitions of new \textit{program relations}. Rules examine and populate an implicit unary program relation $p()$, and stratification conditions are more fine-grained to account for this. For instance, if $p()$ were made explicit, then the reachability rule would be written:
\code{
p(Reach(s)) :- p(Init(s)); p(Reach(s')), p(Trans(s', \dct, s))}.

Static semantics are described through \textit{conforms clauses}; such a clause has the form: $\mathbf{conforms} \ body$.
Where $body$ is any expression that could appear on the right-hand side of a standard rule. A DSL program conforms to its domain if every conforms clause is provable, i.e. conforms clauses are conjunctive. 

The conforms clause in line \exIConformsS~requires a FSM to have at least one initial state. The remaining clauses encode a convention that states and events are ``defined'' by introducing them into the program relation $p()$. Initial states and transitions should only refer to these ``defined'' values. For instance, it is expected that:
$\forall x. \ p(Init(State(x))) \Rightarrow p(State(x))$.
The other conforms clauses check these properties using double-negations. Line \exIInitDef~is read as: \textit{``There should not exist an $x$ where $p(Init(State(x))$ and not $p(State(x))$.''}
The syntax
$\{ head \ | \ body \}$ stands for the set of values produced by the rule $head \head body$, and the $\mathbf{no}$ operator tests if this set is empty. This syntactic sugar allows negations to be nested within a \frm~rule, though negations must still be stratified. Finally, the syntax $x \ \mathbf{is} \ T$ is shorthand for $p(x), x : T$, meaning $x$ holds in $p()$ and has type $T$. This syntax allows variables to range over elements of $p()$ based on their type and helps to avoid long patterns of the form $f(\dc,\ldots,\dc,x,\dc,\ldots,\dc)$.

\textbf{Related work.}~~A number of LP languages support \textit{integrity constraints} (ICs) through rules of the form \textit{false :- body}. A program able to prove the body of an integrity constraint is inconsistent. One could easily envision replacing conforms clauses with ICs. However, there is an important difference. In our approach, the inability to prove conformance is not a logical inconsistency, and this fact allows for more flexible forms of domain composition. The semantics of ICs is rigid and so composing two sets of ICs means all ICs must hold in the composition or the program is logically inconsistent.

\subsection{Models}
\label{sec:models}
\textit{Models} define DSL programs. The contents of a model is a set of well-typed ground facts constructed using the new-kind constructors of the model's domain. Semantically, a model is a logic program formed by composing its domain with the set of ground facts enumerated in the model.
\begin{figure}[t]
\begin{tabular}{@{\hspace{-.04in}}l@{\hspace{-0.1in}} l}
\begin{minipage}{0.5\textwidth}
\begin{FormulaCode}
:model OneStateMach of NonDetFSM
:{
:   State(1).
:   Event("foo").
:   Init(State(1)).
:   Trans(State(1), Event("foo"), 
:         State(1)).
:} 
\end{FormulaCode}
\end{minipage}&
\begin{minipage}{0.485\textwidth}
\vspace{0.025in}
\begin{FormulaCode}
!9
:model TwoStateMach of NonDetFSM {
:   s1 is State(1). 
:   s2 is State(2).
:   eFoo is Event("foo").
:   Init(s1).
:   Trans(s1, eFoo, s2).
:   Trans(s2, eFoo, s2). 
:} 
\end{FormulaCode}
\end{minipage}
\end{tabular}\\
\vspace{-0.1in}
\begin{FormulaCode}
!17
:model BadMach of NonDetFSM { State(1). Init(State(100)). Event("Bar"). } 
\end{FormulaCode}
\caption{Several FSM programs encapsulated within models.}
\label{fig:MachModels}
\end{figure}
Figure \ref{fig:MachModels} shows several examples of FSM models. The \textit{OneStateMach} model defines a FSM consisting of a state with ID 1, an event named ``Foo'', marks 1 as the initial state, and introduces a self-transition on 1 triggered by ``Foo''. Composing these facts with the \textit{NonDetFSM} domain produces a program where all conforms clauses are provable, and so the model is statically correct. The \textit{BadMach} model does not conform, because the initial state (i.e. $State(100)$) is  not an element of $p()$. It violates the conforms clause in line \exIInitDef~of Figure \ref{fig:NonDetFSM}. 

The \textit{TwoStateMach} model uses \textit{symbolic constants}, e.g. $s1$ and $eFoo$, to reuse program expressions.  Symbolic constants are defined in models using the form:
$c  \ \mathbf{is} \ F(t_1,\ldots,t_n)$. Such a definition
introduces the fact $p(F(t_1,\ldots,t_n))$ and defines a module-level constant $c$ that evaluates to $F(t_1,\ldots,t_n)$. Symbolic constants enable snippets of programs to be shared, sometimes leading to exponentially succinct models. The order of definitions does not matter, though every symbolic constant must be defined and definitions must be acyclic.

\section{Domain and Model Composition}
\subsection{Symbol Tables}
Composition is by unioning module definitions. The mechanics of composition depends on symbol definitions and their organization into symbol tables.
\begin{table}[t]
\begin{tabular}{l l l}
\textbf{Symbol kind} & \textbf{Purpose} & \textbf{Example Introduction}\\
\hline ($\eta$) New-kind constructor (arity $> 0$)  & DSL syntax & \code{F ::= new ($\ldots$)}.\\
($\eta$) New-kind constructor (arity $= 0$)  & DSL syntax & 
\code{Nil}, as in \code{U ::= \{ Nil \}. }\\
($\delta$) Derived-kind constructor (arity $> 0$)  & Judgments & \code{G ::= ($\ldots$)}.\\
($\delta$) Dervied-kind constructor (arity $= 0$)  & Judgments  & 
\code{q}, as in \code{q :- F(x).}\\
($\mu$) Union type  & Syntax and judgments  & 
\code{U}, as in \code{U ::= F + \{ 1 \}.}\\
($\nu$) Variable  & Rules & 
\code{x}, as in \code{q :- F(x).}\\
($\sigma$) Symbolic constant  & Alias model expressions & 
\code{c}, as in \code{c is s(s(0)).}\\
\hline
\end{tabular}
\caption{The kinds of user-introduced symbols.}
\label{tab:kinds}
\end{table}    
Table \ref{tab:kinds} lists the kinds of user-defined symbols along with example introductions. Constructors that may appear in a DSL's syntax are called \textit{new-kind} constructors, whereas those only permitted in judgments are called \textit{derived-kind} constructors. Constants are treated as nullary constructors, and derived-kind constants are introduced by using them on the LHS of rules (and can behave like propositions). A standard idiom for declaring a derived-kind constant $q$ is by the tautology $q \head q$. Symbols can only have one kind. For instance, in the rule $x \head f(x)$ the symbol $x$ appears both as a variable and a derived-kind constant, which causes a compile-time error. Built-in symbols such as $1$ and ``foo'' are new-kind constants, and built-in data types such as $I\!nteger$ and $String$ are union types. 

Every module has a symbol table recording all definitions. Let \symbols~be the (infinite) set of all possible symbols, $\kinds \dtb \{ \eta, \delta, \mu, \nu, \sigma  \}$ be the set of symbol kinds, and \terms~be the set of all terms that can be constructed from \symbols. A symbol table $table$ is a partial function from symbols to triples, $table: \symbols \nrightarrow \kinds \times \naturals \times 2^{\terms}$. If $table(s) = (k, n, T)$ then symbol $s$ is defined in $table$ to have kind $k$, arity $n$, and $T$ is the set of well-typed terms that $s$ can represent. Here are some examples:
\[
\begin{array}{c}
table(1) = (\eta, 0, \{1\}), \quad table(x) = (\nu, 0, \terms), \quad table(Integer)= (\mu, 0, \{\ldots, -1, 0, 1, \ldots\}),\\
table(State) = (\eta, 1, \{\ldots,State(-1), State(0), State(1), \ldots\}).
\end{array}
\]
A nullary constructor can only construct a single well-typed term. For non-nullary constructors and union types, $T$ is the set of all well-typed terms obeying their type definitions.   Variables can be substituted for any term. For symbolic constants, $T$ estimates the evaluation of the constants. The composition of symbol tables $table_1$ and $table_2$ is defined as follows. For each $s \in dom \ table_1 \cup dom \ table_2$:
\begin{displaymath}
(table_1 \oplus table_2)(s) \mapsto
\left\{\begin{array}{l}
table_1(s) \ \textrm{if} \
s \in dom \ table_1 - dom \ table_2,\\
table_2(s) \ \textrm{if} \
s \in dom \ table_2 - dom \ table_1,\\
(k, n, T) \ \ \textrm{if} \
(k, n, T) = table_1(s) = table_2(s),\\
\bot \ \ \ \ \ \ \ \ \ \ \textrm{otherwise}.
\end{array}\right.
\end{displaymath}
The composition of tables is legal if no symbol is mapped to $\bot$. This definition allows the same type to be defined in syntactically different ways in several modules. As long as these definitions are semantically equivalent, then the modules can be composed. This flexibility is implemented in \frm.       

\subsubsection{Qualified Symbols and Name Resolution}
Returning to the set of symbols \symbols, we add a free associative operator `.' for constructing qualified symbols. If $x, y \in \symbols$, then $x.y \in \symbols$ and $
(x.y).z = x.(y.z)$. Some examples of qualified symbols are: 
{\small\textit{\begin{center}
$(\delta)$ MyModule.conforms, 
$(\eta)$ Left.State, $(\sigma)$ MyModel.eFoo,
$(\eta)$ In.Left.Trans
\end{center}}}
\noindent Compared to other languages, our module system makes extensive use of qualifiers, so it is important to provide a succinct name resolution strategy. Our strategy is based on \textit{qualifier embeddings}. Let $\vec{q}.s$ be a sequence of atomic symbols $q_1. \ \ldots \ .q_n$ qualifying the atomic symbol $s$. Let $\epsilon$ be the empty sequence of qualifiers, e.g. $\epsilon.s = s$. \textit{A sequence $\vec{p}$ is embedded in $\vec{q}$}, written $\vec{p} \sqsubseteq \vec{q}$, if $\vec{p} = \epsilon$ or there is a monotone function $\iota$ from $\vec{p}$-indices to $\vec{q}$-indices such that $p_1 = q_{\iota(1)}, \ldots, p_{|p|} = q_{\iota(|p|)}$. For example: $b_1.b_2 \sqsubseteq a_1.b_1.a_2.b_2$. The function $lookup(table, \vec{r},\vec{p}.s)$ returns the shortest symbol from $table$ that begins with $\vec{r}$ and embeds $\vec{p}$. Specifically,  
$lookup(table, \vec{r}, \vec{p}.s) \mapsto \vec{r}\vec{q}.s$
if $\vec{p} \sqsubseteq \vec{q}$ and $\vec{r}\vec{q}.s \in dom \ table$. And, for every other $\vec{q'} \neq \vec{q}$ if $\vec{r}\vec{q'}.s \in dom \ table$ then $|\vec{q'}| > |\vec{q}|$. If no such $\vec{q}$ exists then $lookup(table, \vec{p}.s, \vec{r}) \mapsto \bot$. Our $lookup()$ operation searches for symbols in a more general manner than found in other languages. Consequentially, qualified symbols are mostly invisible to the user, even though our composition operators make heavy use of qualifiers. When qualifiers cannot be avoided, the most meaningful qualifiers can be used to disambiguate the symbol. Table \ref{tab:lookup} illustrates the results of $lookup()$ on a sample table.

\begin{table}
\begin{tabular}{l l l}
Look-up operation & Result & Explanation\\
\hline
$lookup(table, \epsilon, f)$ &
$f$ & Symbol with shortest qualifier.\\
$lookup(table, A, A.f)$ &
$A.A.f$ & Symbol with shortest qualifier.\\
$lookup(table, A, A.A.f)$ &
$\bot$ & No such symbol.\\
$lookup(table, \epsilon, B.g)$ &
$\bot$ & Ambiguous, no unique embedding.\\
$lookup(table, \epsilon, B.C.g)$ &
$A.B.C.g$ & Only compatible symbol.\\
$lookup(table, B, C.g)$ &
$\bot$ & No such symbol.\\
\hline
\end{tabular}
\caption{Results of $lookup()$ on a table with the symbols: {\small\textit{f, A.f, A.A.f, A.B.C.g, A.C.B.g}}.}
\label{tab:lookup}
\end{table}

\subsection{Domain Composition}
\label{sec:domcomp}

A domain $D \dtb \langle table, rules \rangle$ consists of a symbol table and set of rules. As a design decision, the derived-kind constants of a domain $D$ are protected by the qualifier $D$. For instance, the rule $q \head q$ actually introduces a symbol $D.q$ into $table$. Every domain has a constant $D.con\!f\!orms$ that is provable if and only if all conforms clauses are provable. The `,' operation merges two domains:
$D_1, D_2 \dtb \langle table_1 \oplus table_2, rules_1 \cup rules_2 \rangle$.
The result is well-defined if $table_1 \oplus table_2$ does not map a symbol to $\bot$, and if the $rules_1 \cup rules_2$ obey stratification conditions. The $includes$ and $extends$ operators allow a domain $D'$ to import a set of domains. 
\begin{displaymath}
\begin{array}{l}
\mathbf{domain} \ D' \ \mathbf{includes} \ D_1,\ldots,D_n \  \{ \ldots \}. \quad
\mathbf{domain} \ D' \ \mathbf{extends} \ D_1,\ldots,D_n \  \{ \ldots \}.
\end{array}
\end{displaymath}
In both cases, $D'$ is treated as if it contains the merged domain $D_1,\ldots,D_n$. The $extends$ operation adds the additional conforms clause to $D'$:
\begin{displaymath}
\textbf{conforms} \ D_1.con\!f\!orms,\ldots,D_n.con\!f\!orms.
\end{displaymath}
The $extends$ operation gives a standard mechanism to conjunct conforms clauses, whereas the $includes$ operation allows the user to define their own composite conformance using the derived-kind constants $D_1.con\!f\!orms$,$\ldots$, $D_n.con\!f\!orms$.

\begin{figure}
\begin{FormulaCode}
:domain DetFSMWithActions extends NonDetFSM, Actions
:{
:   ActMap ::= fun (state: State -> actionName: String).
:   conforms count({ s | Init(s) }) = 1.
:   conforms no { s | Trans(s, e, s'), Trans(s, e, s''), s' != s'' }.
:   conforms no { s | s is State, no Reach(s) }. 
:   conforms no { a | ActMap(_, a), no ActDecl(a, _) }. 
:}
\end{FormulaCode}
\caption{Deterministic FSMs with an action language defined via domain composition.}
\label{fig:DetFSM}
\end{figure}

Figure \ref{fig:DetFSM} uses the \textit{merge} and \textit{extends} operators to define a composite DSL of deterministic FSMs with an action language. The action language (see \ref{app:ActLang}) introduces \textit{state variables} and \textit{actions}, which are sequential programs that update state variables. The static semantics of the $Actions$ domain includes rules for static type-checking of action bodies. The composite DSL introduces an $ActM\!ap$ constructor relating a state to an action that should be executed upon entering that state. The \textbf{fun} keyword is like \textbf{new}, but requires $ActMap$ to be functional (i.e. for all $x$ there is at most one $y$ s.t. $p(ActMap(x, y))$). Finally, the composite language conjoins additional conforms clauses: (1) There is at most one  initial state. (2) The transition table of every state is deterministic. (3) Every state is reachable. (4) Every action named in $ActMap$ has been declared.

\begin{figure}[t]
\begin{FormulaCode}
:model CntrActions of Actions
:{
:  VarDecl("X", INT).
:  ActDecl("ZeroX", Asn("X", 0)).
:  ActDecl("IncX", Asn("X", BnApp(ADD, Var("X"), 1))).    
:}
:model CntrMach of DetFSMWithActions includes TwoStateMach, CntrActions
:{ ActMap(s1, "ZeroX"). ActMap(s2, "IncX"). }  
\end{FormulaCode}
\caption{A counter machine formed by model composition.}
\label{fig:IncMach}
\end{figure}

\subsection{Model Composition}
\label{sec:modelcomp}

A model $M \dtb \langle table, f\!acts \rangle$ is a symbol table and set of facts. $M$ inherits the symbol table of its domain and contains a definition for each symbolic constant $c$ under the fully qualified name $M.c$. Models compose similarly to domains, by composing their symbol tables and unioning fact sets. The $CntrMach$ model in Figure \ref{fig:IncMach} describes a counter machine that increments the state variable ``X'' on every ``Foo'' event. It is defined by composing the $CntrActions$ model with the previous $TwoStateMach$ model. $CntrActions$ declares an integer variable named ``X'', an action named ``ZeroX'' for setting ``X'' to zero, and an action named ``IncX'' for incrementing ``X''. $CntrMach$ assigns state $s1$ to run action ``ZeroX'' and state $s2$ to run action ``IncX''.

\section{Renaming and Transforms}
Transforms are functions from models to models. They are defined with ADTs, strongly-typed logic programs, and a module-level operator called \textit{renaming}.
\subsection{Renaming}
\label{sec:renaming}
The renaming operator ``$::$'' produces a new module $x\!::\!M$  that is identical to $M$, except that every occurrence of symbol $s$ in $M$ becomes $x.s$ in $x\!::\!M$. As a design decision, variables and new-kind constants retain their original names. The renaming operator allows the easy construction of distinguished copies of modules. For instance, renaming can be used to construct a DSL representing two distinct FSMs running in parallel. Figure \ref{fig:Parallel} shows how this DSL can be constructed. The \textit{ParallelFSMs} domain is a product domain containing two copies of \textit{DetFSMWithActions} under the renamings \textit{left} and \textit{right}. Symmetrically, the renaming operator can be applied to models. The \textit{ParallelCntrs} model contains two renamed copies of \textit{CntrMach}, thereby creating a valid \textit{ParallelFSMs} model. Table \ref{tab:sympara} lists the symbol table of this composite model.

\subsection{Transforms}
\label{sec:transforms}
\begin{figure}[t]
\begin{FormulaCode}
:domain ParallelFSMs extends left::DetFSMWithActions, right::DetFSMWithActions
:{}
:model ParallelCntrs of ParallelFSMs includes left::CntrMach, right::CntrMach 
:{}
\end{FormulaCode}
\caption{Using renaming to create a DSL for two FSMs running in parallel.}
\label{fig:Parallel}
\end{figure}

Transforms use renaming to label their inputs and outputs. A transform has the shape:
\[
\mathbf{transform} \ T \ (x_1 :: D_1,\ldots, x_m :: D_m) \ \mathbf{returns} \
(y_1 :: E_1,\ldots, y_n :: E_n) \ \{ \ldots \}.
\]
where $D_i$ and $E_j$ are domains and the labels $x_1,\ldots, x_m, y_1, \ldots, y_n$ are distinct. The transform module $T$ is a composition of its body with the renamed domains in its signature. The signature also indicates that  $x_1 :: D_1, \ldots, x_m:: D_m$ are \textit{input domains} and $y_1::E_1,\ldots, y_n:: E_n$ are \textit{output domains}.

\begin{figure}[t]
\begin{FormulaCode}
:transform Prune (in:: NonDetFSM) returns (out:: NonDetFSM)
:{
ReqConf:   requires in.conforms.
EnsConf:   ensures out.conforms.
eventCopy:   out.Event(n) :- in.Event(n).
stateCopy:   out.State(x) :- in.Reach(State(x)).
initCopy:   out.Init(s)  :- in.Init(s).
:   out.Trans(s, e, s') :- in.Trans(s, e, s'), in.Reach(s), in.Reach(s').
:} 
\end{FormulaCode}
\caption{A transform that prunes dead states.}
\label{fig:Prune}
\end{figure}

Figure \ref{fig:Prune} shows a transform that takes an FSM as input and outputs an equivalent FSM where unreachable states and transitions have been pruned away. Notice the use of renamed constructors to distinguish between input and output values. The first rule (line \eventCopy) copies all event declarations from the input to the output. The next rule copies only reachable states. The expression $in.Reach(State(x))$ only requires one qualification, even though $in.Reach()$ must have arguments of type $in.State$. This rule is legal because name resolution first uses the qualifier on the outer constructor to resolve the name of an inner constructor (e.g. $lookup(table, in, State)$). If this resolution fails, then it is retried without the outer qualifier.

\subsection{Inferred Term Rewrites}
\label{sec:rewrites}

Every initial state is by reachable, so the third rule copies all initial states (line \initCopy). This rule looks trivial, but closer inspection reveals a potential problem. Variable $s$ in the RHS must have type $in.State$, but $s$ in the LHS must have type $out.State$. Because these types are disjoint, the compiler should reject this rule as badly typed. Instead, the compiler infers the user's intent to convert values of the form $in.State(a)$ to $out.State(a)$. The rule inferred by the compiler is actually:
\code{out.Init(\relabel{in}{out}(s)) :- in.Init(s)}. The \textit{relabeling function} $\relabel{in}{out}$ rewrites terms by replacing the \textit{in} qualifier with the \textit{out} qualifier. Relabeling functions are recursive; they rewrite arbitrarily deep terms. In our experience, inferred rewrites are essential for making renaming operator practical. 

Due to space limitations we only define the inference problem, but not the requisite algorithms. Let $x$ be a variable occurring in the RHS of a rule and the set $T_{rhs}$ contain (at least) all the values $x$ takes when the body of the rule is satisfied. Also, suppose $x$ occurs somewhere in the LHS and $T_{lhs}$ is the set of all values that $x$ is allowed to take in the LHS. Then $x$ is well-typed if there exists a \textit{unique} relabeling function \relabel{\vec{p}}{\vec{q}} s.t. $\relabel{\vec{p}}{\vec{q}}(T_{rhs}) \subseteq T_{lhs}$. The relabeling of a new-kind constant is itself. Otherwise:
\[
\relabel{\vec{p}}{\vec{q}}(\vec{p}\vec{u}.f(t_1, \ldots, t_n)) \dtb
\vec{q}\vec{u}.f(\relabel{\vec{p}}{\vec{q}}(t_1), \ldots, \relabel{\vec{p}}{\vec{q}}(t_n)).
\]
Deciding if such a relabeling exists is non-trivial and requires complex type inference algorithms. These have been implemented in \frm.

\subsection{Execution, Contracts and Composition}
\label{sec:exe}
A transform $T$ is applied to a sequence of models $M_1,\ldots,M_m$ defined over input domains $D_1,\ldots, D_m$. The mechanics of application are as follows: First, $T$ is composed with the renamed models $x_1::M_1, \ldots, x_m::M_m$ and the resulting logic program is executed. Next, the $j^{th}$ output model $N_j$ is constructed by collecting all values $t$ such that $p(t)$ holds and $t = y_j\vec{u}.f(\ldots)$. Finally, the renaming $y_j$ is removed from these values yielding an output model purely in the output domain $E_j$. In symbols:
\[
N_j \dtb \{ \ \relabel{y_j}{\epsilon}(t) \ | \
p(t) \ \textrm{and} \
t = y_j\vec{u}.f(\ldots) \ \textrm{and} \
kind(\vec{u}.f) = \eta
\}.
\]
As usual, only new-kind constructors can appear in output models, hence the extra constraint on the kind of the constructor.

\textit{Contracts} appear in many programming languages for clearly specifying the intent of methods / functions \cite{NienaltowskiMO09,BarnettS03}. They can be checked at run-time or compile-time tools can attempt to prove their validity.  Our contracts allow users to specify required properties of input models, and properties ensured by the transform when all requirements are met. A clause of the form $\mathbf{requires} \ body$ states a required property. A clause of the form $\mathbf{ensures} \ body$ states an ensured property. For example, line \ReqConf~of \textit{Prune} requires the input model to conform to its domain. Line \EnsConf~guarantees output conformance when the input conforms. As with conforms clauses, transform contracts are conjunctive, and each transform $T$ has derived-kind constants \textit{T.requires} and \textit{T.ensures} that are provable when all requires and ensures clauses are satisfied. A transform contract is a claim that for every application $T(M_1,\ldots,M_m)$ then $p(T.requires) \Rightarrow p(T.ensures)$. Compile-time verification is supported by \frm, but outside the scope of this paper. 

\begin{figure}[t]
\begin{FormulaCode}
:transform system PruneAndParallelize (in1:: NonDetFSM, in2:: NonDetFSM)
:returns (out:: ParallelFSMs)
:{
:   prune1 = Prune(in1).
:   prune2 = Prune(in2).
:   out    = Parallelize(prune1, prune2).
:}
\end{FormulaCode}
\caption{Sequential composition of transforms.}
\label{fig:seq}
\end{figure}

Finally, transforms can be sequentially composed by listing a series of oriented equations as shown in Figure \ref{fig:seq}. The \textit{PruneAndParallelize} composite transform takes two FSMs as inputs, prunes the FSMSs, and then combines them into a single \textit{ParallelFSMs} model. The RHS of each equation is a transform application, and the LHS is a sequence of variables that will be bound to the outputs produced by the application. Such a composite is executed by running each constituent transform in dependency-order, while applying renaming and un-renaming at the boundaries of each step. Transform systems can also be sequentially composed within transform systems.

\subsection{Related Work}
\textit{Prolog} specifically \cite{JouaultB06} and LP generally \cite{HorvathBRV10} have been recognized as useful for \textit{model transformations}. In the former case, Prolog is employed as a behind-the-scenes execution engine, so it was not extended with modules. In the latter case, Prolog-style semantics served as inspiration for the \textit{VIATRA2} language, which is not a LP language. However, model transformation languages such as \textit{VIATRA2} do commonly support various forms of sequential composition \cite{BisztrayHE09,BoronatHM09}, giving further motivation to include it in a module system. LP with constraints has also been used to reason about DSL specifications by translation to LP without extending its module system \cite{CabotCR07}.  

DSLs have a long history in the software modeling community.  The tools and techniques of the modeling community have been heavily influenced by the \textit{Unified Modeling Language} (UML), the \textit{Object Constraint Language}, and the concept of \textit{metamodeling}. However, the resulting amalgam is difficult to formalize \cite{BoronatM10} and has a complex and underdeveloped module system \cite{DingelDZ08}. By adding a module system for DSLs directly on top of LP, we inherit the well-understood semantics of LP and avoid problems associated with other notations.

\section{Discussion and Future Work}
In this paper we presented a complete module system on top of LP for the construction, composition, and reuse of DSLs. Our modules, composition operators, and language extensions were designed to be simple yet synergistic. For instance, inferred term rewrites are a general concept and can also be utilized in all modules. Though our running example illustrated a simple compiler, the same mechanism can also be used to specify the elementary steps of transition systems. This use-case is important for dynamic semantics. Our module system evolved over a period of several years, and was informed by both academic and industrial applications of LP for DSL design. For example, we found a contract language to be essential for documenting the intent of modules in an actionable form. 

There are still some important ways in which our module system could be extended. One natural desire is a mechanism to make transforms polymorphic on domains. Another is to package the set of transforms defining an abstract transition system into a single module. The contract language for such a package is also an interesting design problem and would most likely fit well with some form of temporal logic. In our opinion more experimentation and use-cases are required before the right combination of extensions becomes clear. We are currently conducting these experiments while applying \frm~to industrial examples. The examples presented in this paper along with a binary version of \frm~can be found at
\url{http://research.microsoft.com/en-us/um/redmond/projects/formula/ICLP2014.html}. The binary version is compiled for Windows machines. The source code for \frm~and along with its implementation of this module system can be found at \url{formula.codeplex.com}.

\clearpage
\appendix
\section{Action Language}
\label{app:ActLang}
\begin{FormulaCode}
:domain Actions
:{
:   //// Declarations
:   VarDecl ::= fun (id: String -> type: { BOOL, INT }).
:   ActDecl ::= fun (id: String -> action: any Action).
:   
:   //// Action language (expressions)
:   BoolOp ::= { NOT, AND, OR }.
:   IntOp  ::= { NEG, ADD, SUB, MUL, DIV }.
:   CmpOp  ::= { LT, LE, GT, GE, EQ, NEQ }.
:   
:   Var   ::= new (id: String).
:   UnApp ::= new (op: { NEG, NOT }, arg1: any Expr).  
:   BnApp ::= new (op: { ADD, SUB, MUL, DIV, AND, OR } + CmpOp, 
:                  arg1: any Expr, arg2: any Expr).
:   Expr  ::= Var + UnApp + BnApp + Boolean + Integer.      
:   
:   //// Action language (statements)
:   Asn ::= new (var: String, expr: any Expr).
:   ITE ::= new (cond: any Expr, true: any Action, false: any Action).
:   Seq ::= new (act1: any Action, act2: any Action).
:   Action ::= Asn + ITE + Seq + { NOP }.
:   
:   //// Static typing
:   Sub ::= (Action + Expr).
:   Sub(e)                    :- ActDecl(_, e); 
:                                Sub(UnApp(_, e));
:                                Sub(Asn(_, e)).
:   Sub(e), Sub(e')           :- Sub(BnApp(_, e, e')); 
:                                Sub(Seq(e, e')).
:   Sub(e), Sub(e'), Sub(e'') :- Sub(ITE(e, e', e'')).
:   
:   TypeJudge ::= (Action + Expr, { BOOL, INT, ANY }).
:   
:   TypeJudge(e, INT) :- 
:      Sub(e), e : Integer;
:      Sub(e), e = Var(n), VarDecl(n, INT);
:      Sub(e), e = UnApp(op, e'), op : IntOp, TypeJudge(e', INT);
:      Sub(e), e = Asn(n, e'), VarDecl(n, INT), TypeJudge(e', INT);
:      Sub(e), e = BnApp(op, e', e''), op : IntOp, 
:         TypeJudge(e', INT), TypeJudge(e'', INT).
:                        
:   TypeJudge(e, BOOL) :- 
:      Sub(e), e : Boolean;
:      Sub(e), e = Var(n), VarDecl(n, BOOL);
:      Sub(e), e = UnApp(op, e'), op : BoolOp, TypeJudge(e', BOOL);
:      Sub(e), e = Asn(n, e'), VarDecl(n, BOOL), TypeJudge(e', BOOL).   
:      Sub(e), e = BnApp(op, e', e''), op : BoolOp, 
:         TypeJudge(e', BOOL), TypeJudge(e'', BOOL);
:      Sub(e), e = BnApp(op, e', e''), op : CmpOp, 
:         TypeJudge(e', t), TypeJudge(e'', t).
:                         
:   TypeJudge(e, ANY) :-  
:      Sub(e), e = NOP;
:      Sub(e), e = Seq(e', e''), TypeJudge(e', _), TypeJudge(e'', _);
:      Sub(e), e = ITE(e', e'', e'''), TypeJudge(e', BOOL), 
:         TypeJudge(e'', _), TypeJudge(e''', _).
:                                           
:   conforms no { e | Sub(e), no { t | TypeJudge(e, t) } }.                                                                          
:}
\end{FormulaCode}

\clearpage
\section{Example Symbol Table}
\label{app:ExSymTab}
\begin{table}[h]
{\footnotesize\texttt{
\begin{tabular}{r | l | c || r | l | c}
Qualifiers     &     Name      & Kind, Arity & Qualifiers     &     Name      & Kind, Arity\\[-6pt]
\multicolumn{6}{l}{\rule{\textwidth}{.1pt}}\\
                   &      ADD      &   $\eta$, 0&          left           &     Init      &   $\eta$, 1\\
                         &      AND      &  $\eta$, 0&          left           &     IntOp     &   $\mu$, 0\\
                         &      ANY      &  $\eta$, 0&          left           &     Reach     &   $\delta$, 1\\
                         &     BOOL      &   $\eta$, 0&          left           &      Seq      &   $\eta$, 2\\
                         &      DIV      &  $\eta$, 0&          left           &     State     &   $\eta$, 1\\
                         &      EQ       &   $\eta$, 0&          left           &      Sub      &   $\delta$, 1\\
                         &     FALSE     &  $\eta$, 0&          left           &     Trans     &   $\eta$, 3\\
                         &      GE       &  $\eta$, 0&          left           &   TypeJudge   &   $\delta$, 2\\
                         &      GT       &   $\eta$, 0&          left           &     UnApp     &   $\eta$, 2\\
                         &      INT      &   $\eta$, 0&          left           &      Var      &   $\eta$, 1\\
                         &      LE       &   $\eta$, 0&          left           &    VarDecl    &   $\eta$, 2\\
                         &      LT       &   $\eta$, 0&      left.Actions       &   conforms    &   $\delta$, 0\\
                         &      MUL      &   $\eta$, 0& left.DetFSMWithActions  &   conforms    &   $\delta$, 0\\
                         &      NEG      &   $\eta$, 0& left.MachTwoState       &      eFoo     &   $\sigma$, 0\\
                         &      NEQ      &   $\eta$, 0& left.MachTwoState       &      s1       &   $\sigma$, 0\\
                         &      NOP      &   $\eta$, 0& left.MachTwoState       &      s2       &   $\sigma$, 0\\
                         &      NOT      &   $\eta$, 0&     left.NonDetFSM      &   conforms    &   $\delta$, 0\\
                         &      OR       &   $\eta$, 0&          right          &    ActDecl    &   $\eta$, 2\\
                         &      SUB      &   $\eta$, 0&          right          &    ActMap     &   $\eta$, 2\\
                         &     TRUE      &   $\eta$, 0&          right          &    Action     &   $\mu$, 0	\\
                         &       a       &   $\nu$, 0&          right          &      Asn      &   $\eta$, 2\\
                         &       e       &   $\nu$, 0&          right          &     BnApp     &   $\eta$, 3\\
                         &       e'      &   $\nu$, 0&          right          &    BoolOp     &   $\mu$, 0\\
                         &       e''     &   $\nu$, 0&          right          &     CmpOp     &   $\mu$, 0\\
                         &       e'''    &   $\nu$, 0&          right          &     Event     &   $\eta$, 1\\
                         &       i       &   $\nu$, 0&          right          &     Expr      &   $\mu$, 0\\ 
                         &       n       &   $\nu$, 0&          right          &      ITE      &   $\eta$, 3\\
                         &       op      &   $\nu$, 0&          right          &     Init      &   $\eta$, 1\\
                         &       s       &   $\nu$, 0&          right          &     IntOp     &   $\mu$, 0\\
                         &       s'      &   $\nu$, 0&          right          &     Reach     &   $\delta$, 1\\
                         &       s''     &   $\nu$, 0&          right          &      Seq      &   $\eta$, 2\\
                         &       t       &   $\nu$, 0&          right          &     State     &   $\eta$, 1\\
      ParallelFSMs       &   conforms    &   $\delta$, 0&          right          &      Sub      &   $\delta$, 1\\
          left           &    ActDecl    &   $\eta$, 2&          right          &     Trans     &   $\eta$, 3\\
          left           &    ActMap     &   $\eta$, 2&          right          &   TypeJudge   &   $\delta$, 2\\
          left           &    Action     &    $\mu$, 2&          right          &     UnApp     &   $\eta$, 2\\
          left           &      Asn      &   $\eta$, 2&          right          &      Var      &   $\eta$, 1\\
          left           &     BnApp     &   $\eta$, 3&          right          &    VarDecl    &   $\eta$, 2\\
          left           &    BoolOp     &   $\mu$, 0&      right.Actions      &   conforms    &   $\delta$, 0\\
          left           &     CmpOp     &   $\mu$, 0& right.DetFSMWithActions &   conforms    &   $\delta$, 0\\
          left           &     Event     &   $\eta$, 1& right.MachTwoState      &      eFoo     &   $\sigma$, 0\\
          left           &     Expr      &   $\mu$, 0& right.MachTwoState      &      s1       &   $\sigma$, 0\\
          left           &      ITE      &   $\eta$, 3& right.MachTwoState      &      s2       &   $\sigma$, 0\\
                         &               &                                      &     right.NonDetFSM     &   conforms    &   $\delta$, 0\\[-6pt]
\hline                         
\end{tabular}}}
\caption{Symbol table of composite model \textit{ParallelCntrs} in Figure \ref{fig:Parallel}.}
\label{tab:sympara}
\end{table}

\clearpage
\bibliographystyle{acmtrans}
\bibliography{ModSystem}

\end{document}